# Post-Lecture Interactive Environments for Conceptual Learning: A Randomized Comparison of Mixed Reality and Tangible Instruction in Undergraduate STEM Education

**Sharmin Akter[1], Mohammad Abu Nasir Rakib[2], Eshwara Prasad Sridhar[3], Somik Biswas[3], Mahmudur Rahman[3], and Md Rassel Raihan[1]**

[1] Department of Mechanical and Aerospace Engineering, The University of Texas at Arlington, Arlington, TX, USA
[2] Department of Computer Science and Engineering, The University of Texas at Arlington, Arlington, TX, USA
[3] Department of Industrial, Manufacturing, and Systems Engineering, The University of Texas at Arlington, Arlington, TX, USA

**Corresponding author**
Md Rassel Raihan
Assistant Professor, Department of Mechanical and Aerospace Engineering
The University of Texas at Arlington
Email: mdrassel.raihan@uta.edu

**ABSTRACT**
Developing conceptual understanding in engineering requires learners to connect spatial reasoning with abstract representations, yet lecture-based instruction often provides limited support for this process. Interactive learning environments, including mixed reality (MR) and tangible tools, may help students revisit difficult concepts through action, feedback, and visible system response. This pilot randomized study compared two post-lecture interventions, an immersive MR application and a tangible Engineering Toolkit, with lecture-only instruction in undergraduate solid mechanics. Twenty-four participants completed a baseline assessment, a common lecture, and a post-instruction knowledge test; participants in the interactive conditions also completed usability and learner-experience measures. Learning outcomes were analyzed using ANCOVA with baseline knowledge as a covariate and were supported by normalized learning gains. Instructional modality had a significant effect on post-test performance, $F(2, 20) = 3.60$, $p = .048$, partial $\eta^2 = .263$. Both interactive conditions outperformed lecture-only instruction in planned contrasts. MR showed the highest normalized learning gain ($g = .57$), whereas the tangible toolkit showed higher usability and learner confidence. In the MR condition, virtual reality sickness symptoms measured via the Virtual Reality Sickness Questionnaire remained low before and after the intervention, suggesting that the MR application was well-tolerated by participants. These results suggest that post-lecture interactive environments may support immediate conceptual learning while offering different modality-specific strengths that require larger, time-matched follow-up studies.

**KEYWORDS**
mixed reality, technology-enhanced learning, tangible learning, conceptual understanding, spatial reasoning, engineering education

## 1. Introduction

Interactive and immersive technologies are increasingly used to support conceptual learning, especially in domains where learners must connect abstract representations with spatial reasoning. Digital tools such as virtual reality (VR), augmented reality (AR), and mixed reality (MR) have been shown to improve visualization, engagement, and understanding in engineering and related fields (Kraus et al., 2021; Pàmies-Vilà et al., 2025; Tang et al., 2020).

At the same time, physical manipulatives and hands-on models provide a different form of support by allowing learners to interact directly with systems, which can strengthen mental model formation through embodied experience (Kullberg et al., 2017; Sorby, 2009; Wilson, 2002). Together, these approaches offer promising ways to support conceptual learning beyond traditional lecture-based instruction.

Despite these advances, several limitations remain. First, many studies focus on a single intervention compared with lecture-based instruction, making it difficult to understand how different interactive modalities compare within the same learning context. Second, prior work often emphasizes performance outcomes while giving less attention to learner experience factors such as confidence, self-efficacy, and usability, which are important for adoption and sustained use. Third, relatively few studies examine how interactive tools function as extensions to lecture-based instruction, particularly as post-lecture interventions designed to reinforce and deepen understanding.

To address these gaps, this pilot study investigates two post-lecture interactive learning environments: an immersive Mixed Reality (MR) application and a tangible engineering toolkit. Both are compared with lecture-only instruction in the context of solid mechanics. All participants first completed a common lecture before engaging with condition-specific post-lecture activities, allowing the analysis to focus on the added value of interactive support after shared instruction. The study uses a randomized, baseline-controlled design to evaluate conceptual learning and examine learner self-efficacy and usability. This study addresses the following research questions:

**RQ1:** To what extent do MR and a tangible toolkit improve conceptual understanding compared with lecture-only instruction?

**RQ2:** How do learner confidence and satisfaction differ between MR and tangible interventions?

**RQ3:** How do learners perceive the usability of these interactive modalities?

This work makes three contributions. First, it provides a controlled comparison of immersive and tangible post-lecture learning environments within a shared instructional context. Second, it integrates learning outcomes with usability and learner-experience measures to provide a more complete evaluation of instructional effectiveness. Third, it examines these effects in a spatially demanding STEM domain, offering design-relevant insights for technology-enhanced learning environments in contexts where learners must coordinate visual, physical, and conceptual representations.

## 2. Background

### 2.1 Conceptual challenges in solid mechanics learning

Learning in solid mechanics is difficult because students must coordinate symbolic, graphical, and spatial representations of phenomena that are often not directly visible. Concepts such as stress, strain, bending, and deformation require learners to connect equations and diagrams with internal structural behavior and geometric change in space. This makes conceptual understanding harder than simply applying formulas, since students must interpret relationships among loads, supports, internal forces, and deformation in ways that are often abstract and spatially demanding (Ha & Fang, 2016; Higley et al., 2007; Litzinger et al., 2010; Streveler et al., 2008).

Prior work in engineering education has shown that difficulties in mechanics are closely tied to representational reasoning and spatial ability. Students may be able to reproduce procedures without fully understanding the underlying physical meaning of the problem. In statics and related domains, conceptual understanding improves when students can interpret diagrams, coordinate visual and symbolic information, and reason about structural behavior beyond surface-level problem solving (Johnson-Glauch & Herman, 2019; Litzinger et al., 2010). Spatial visualization is especially important because learners often need to mentally transform objects, inspect hidden relationships, and infer three-dimensional behavior from two-dimensional representations (Alias, 2002; Hsi et al., 1997; Sorby, 2009).

Research also suggests that spatial skills are not fixed and can be improved through targeted instructional support. Meta-analytic and domain-specific work indicates that appropriate learning experiences can strengthen learners' ability to interpret and manipulate spatial information, which in turn can support performance in engineering contexts (Sorby, 2009; Uttal et al., 2013). This is important for mechanics education because many core concepts depend on how well students connect formal representations to physical behavior. As a result, learning environments that make these relationships easier to inspect and manipulate may help students develop stronger conceptual understanding.

### 2.2 Interactive and representational support for learning

A central implication of this literature is that external representations matter for learning. Students do not build understanding only from verbal explanation or symbolic notation. They also rely on diagrams, models, and other representations that help map abstract quantities onto observable or interpretable forms. The DeFT framework and related work on multiple representations suggest that learning can improve when representations complement one another and make key relationships more accessible to learners (Ainsworth, 2006). In spatially demanding domains, this is especially relevant because well-designed representations can reduce the burden of mental visualization and support conceptual interpretation (Hegarty, 2004a).

In engineering education, representational support is valuable because many novice errors arise from weak connections between symbolic procedures and physical meaning. Students often struggle to interpret what diagrams or equations imply about structural response, even when they can complete procedural steps. Visual representations can help guide problem-solving strategies, while opportunities to inspect and manipulate representations can support deeper conceptual reasoning (Johnson-Glauch & Herman, 2019). This makes interactive learning particularly relevant in mechanics, where important features of the problem are often hidden or dynamic rather than directly observable.

Embodied cognition provides an additional rationale for using interactive supports. From this perspective, cognition is shaped not only by abstract mental processing but also by bodily action and perceptual experience (Wilson, 2002). Physical interaction with learning materials may therefore help students ground abstract concepts in more concrete experiences (Byrne et al., 2023; Castro-Alonso et al., 2024). In mechanics, where learners must often reason about force, motion, and structural behavior, interaction may support understanding by linking conceptual ideas to direct perceptual or manipulative experiences (Negahban, 2024). This suggests that the educational value of interactive tools lies not only in novelty or presentation format, but in how they shape learners' engagement with representations (Lichtenberger et al., 2024; McNeil & Fyfe, 2012).

### 2.3 Mixed reality, tangible learning, and research gaps

Mixed reality is one promising way to support this kind of learning. By allowing students to inspect and interact with three-dimensional representations, mixed reality can make difficult spatial relationships easier to explore. Reviews of immersive learning technologies in higher education have reported potential benefits for engagement, visualization, and conceptual learning, particularly in STEM domains where spatial interpretation is important (Jensen & Konradsen, 2018; Radianti et al., 2020a).

In engineering education, mixed reality has been used to support visualization and design-related learning, with evidence suggesting that learners may benefit when they can interact more directly with 3D content rather than rely only on static instructional materials (Koh et al., 2010; Tang et al., 2020). Recent work has continued to highlight the promise of augmented and mixed reality in engineering and technical education, while also noting the need for stronger evidence about the specific conditions under which these tools improve learning (Garzón et al., 2021; Koumpouros, 2024; Pàmies-Vilà et al., 2025). A related but distinct approach involves tangible and physical learning supports. Tangible tools can make mechanics concepts more concrete by allowing students to manipulate physical models and observe structural relationships directly. Prior work has shown that self-constructed and physical representations can support the learning of free-body diagrams, statics, and related topics by helping students externalize and test their reasoning (Davishahl et al., 2020; Mešic et al., 2016; Sadowski & Jankowski, 2021).
These approaches are consistent with the view that direct physical interaction can support conceptual understanding in ways that differ from screen-based or immersive digital tools (Koç & Kanadlı, 2025; Radianti et al., 2020a).
In this sense, mixed reality and tangible supports may both help learning, but through somewhat different representational and interactional pathways(Coban et al., 2022; Kaplan et al., 2021).

Despite this promise, important gaps remain in the literature. Many studies evaluate one intervention in isolation or compare a single tool against traditional instruction (Coban et al., 2022; Radianti et al., 2020). Fewer studies compare two different interactive modalities under a shared instructional baseline, which makes it difficult to understand how different post-lecture supports may influence conceptual learning. In addition, prior work often centers on performance outcomes while giving less attention to learner-centered factors such as usability, confidence, and satisfaction, even though these factors matter for whether an intervention is practical and educationally valuable in classroom settings (Bangor et al., 2008; Lewis & Sauro, 2009).

These gaps matter because post-lecture support is a realistic point of intervention in engineering education. After initial exposure to difficult concepts, students often need opportunities to revisit and reorganize their understanding using richer representational support. Comparing mixed reality and tangible post-lecture tools can therefore help clarify not only whether interactive supports improve conceptual understanding, but also how different forms of interaction shape learner experience. Framed this way, the study has relevance beyond a single topic or device and speaks more broadly to the design of technology-enhanced learning environments for spatially demanding STEM concepts.

## 3. Methodology

### 3.1 Study design and participants

This pilot study used a randomized between-subjects design to examine how post-lecture instructional modality influences conceptual learning in solid mechanics. Three conditions were compared: lecture-only instruction, lecture followed by a Mixed Reality (MR) activity, and lecture followed by a tangible Engineering Toolkit (ET) activity. The design isolates the added value of different post-lecture learning experiences after a common instructional baseline. The instructional topic was beam bending, chosen as a representative solid mechanics concept with high spatial and representational demands. Understanding beam bending requires learners to reason about support reactions, load-position effects, shear force, and bending moment while linking diagrams to physical behavior. These tasks require coordination across multiple representations, making beam bending an appropriate context for examining whether interactive post-lecture supports strengthen conceptual understanding. Twenty-four undergraduate students participated in the study. All had completed Engineering Statics, ensuring a common prerequisite background in foundational mechanics concepts. Participants were recruited through flyers and email announcements at the university. After completing the baseline assessment, participants were randomly assigned to one of the three conditions, with eight participants per group. Those assigned to the MR condition also completed standard safety screening for immersive participation. The study was conducted in a controlled laboratory setting using a standardized researcher script to minimize procedural variation across sessions. Figure 1 summarizes the study sequence, including shared pre-instruction steps and condition-specific post-lecture activities.

### 3.2 Instructional conditions and design considerations

All three conditions targeted the same concepts: support reactions, load-position effects, and interpretation of shear force and bending moment relationships. They differed only in post-lecture engagement, allowing comparison across modalities while holding conceptual content constant. Figure 2 shows the conditions. Both interactive conditions follow shared design principles. First, they reduced unnecessary processing by focusing on a single beam scenario and a limited set of key actions. Second, they supported close coupling between learner actions and system responses. Third, they made abstract relationships more visible through either physical interaction or interactive spatial visualization. These principles were embedded in the activity design rather than presented explicitly.

### 3.3 Common lecture

All participants viewed the same prerecorded lecture on beam bending, introducing core concepts through slides, diagrams, and worked examples. Narration used a neutral text-to-speech voice to reduce variability. Participants could take notes and adjust playback speed, ensuring consistent content while allowing individual pacing. This lecture served as the common baseline.

### 3.4 Lecture-only condition

Participants completed the post-instruction knowledge quiz immediately after the lecture and then ended the session. This condition served as the baseline for evaluating the added value of interactive interventions.

### 3.5 Tangible Engineering Toolkit condition

The ET condition used a physical beam apparatus with a marked beam, removable weights, two support scales, and fixed load positions. Participants completed three guided trials at set positions, followed by self-directed exploration. By repositioning loads and observing changes in support reactions, learners engaged in direct manipulation with immediate perceptual feedback. This setup externalized equilibrium relationships in a concrete reference frame, reducing reliance on mental simulation.

### 3.6 Mixed Reality condition

The MR condition presented the same beam scenario in an immersive environment. Participants adjusted load positions and observed structural responses using hand input or joystick controls with haptic feedback. The interface maintained a stable spatial layout, linking actions to system changes. It was intentionally simple to reduce interface demands while supporting attention to relationships among load position, reactions, and graphical representations.

### 3.7 Procedure

Each session followed the same overall structure. After arrival, participants completed informed consent, a demographic questionnaire, and a baseline knowledge quiz. All participants then viewed the 30-minute common lecture. After the lecture, participants continued according to their assigned condition. In the lecture-only condition, participants completed the post-instruction knowledge quiz and then departed. In the ET condition, participants completed the engineering toolkit experience, followed by the post-instruction knowledge quiz, the System Usability Scale (SUS), and the learner self-efficacy questionnaire, and then concluded the session. In the MR condition, participants first completed a pre-exposure Virtual Reality Sickness Questionnaire (VRSQ), then engaged with the MR activity, completed the post-exposure VRSQ, followed by the post-instruction knowledge quiz, SUS, and learner self-efficacy questionnaire, and then concluded the session. The full procedure is shown in Figure 1. The instructional conditions are illustrated in Figure 2. The procedure was delivered using a standardized script by trained researchers to reduce session-level differences in instruction or facilitation. A reasonable concern in this design is whether differences between groups reflect modality-specific affordances or simply additional learning time after the lecture. In this study, the primary aim was to compare lecture-only instruction with lecture plus interactive extension and then compare two forms of interactive extension under the same conceptual target. Exact time-on-task during the interactive activities was not formally logged; therefore, the resulting interpretation is focused on the educational value of post-lecture interactive support rather than replacement of lecture instruction or pure modality effects.

***Table 1. Summary of instructional conditions***

| **Component** | **Lecture-only** | **Engineering Toolkit (ET)** | **Mixed Reality (MR)** |
|---|---|---|---|
| Post-lecture activity | None | Physical beam manipulation | Immersive virtual interaction |
| Interaction type | Passive | Hands-on physical | Spatial interactive |
| Feedback type | None | Direct physical feedback (scales, weights) | Visual + interactive feedback |
| Representation | Slides/diagrams | Physical beam system | Virtual beam system |
| Guidance | None | Guided trials + exploration | Guided interaction |
| Key affordance | Concept exposure | Embodied interaction | Spatial visualization |

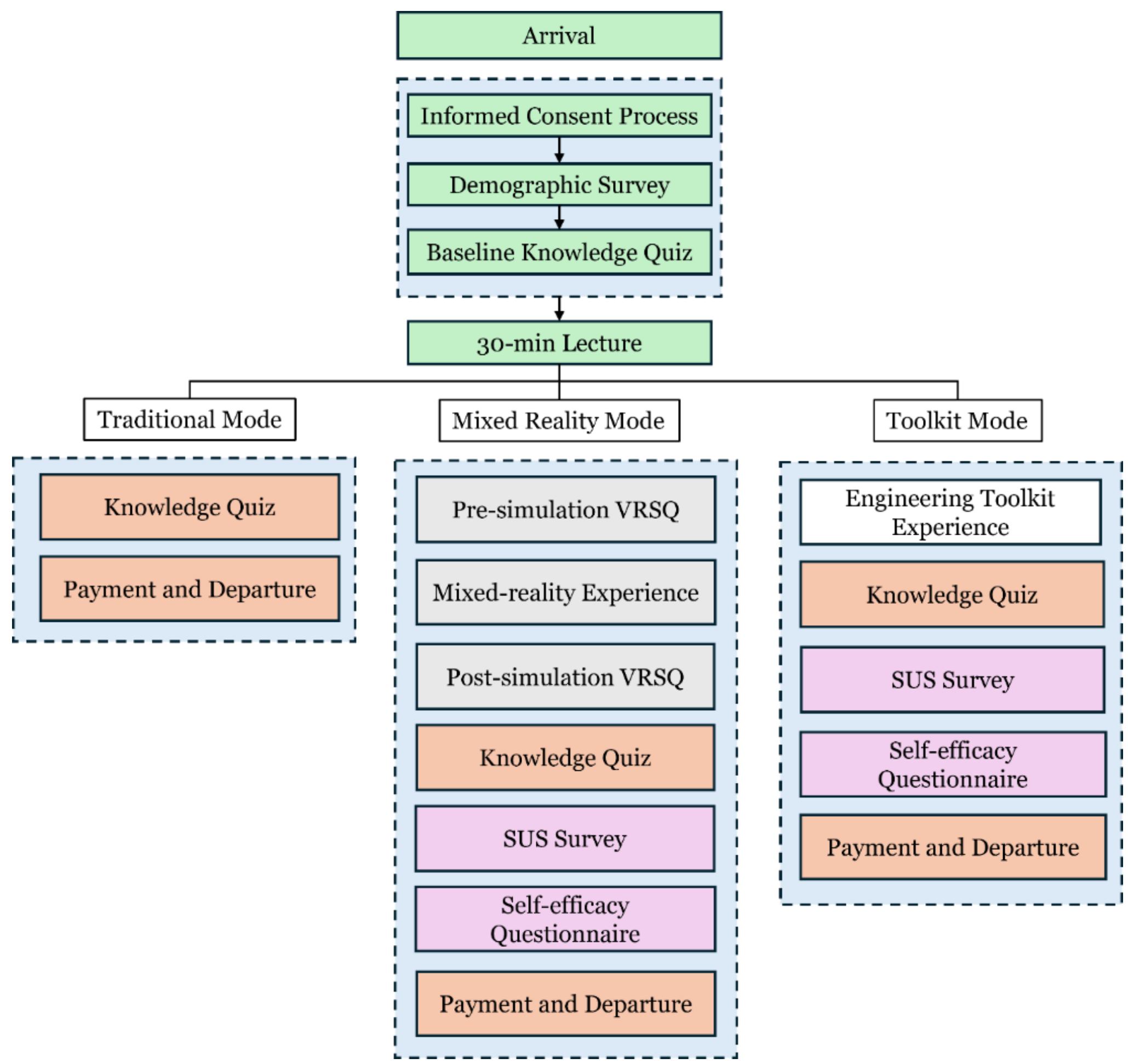

***Figure 1*** *Study procedure for the three instructional conditions.*

## 3.8 Measures

### 3.8.1 Knowledge assessment

Participants completed a baseline quiz before instruction and a post-instruction quiz after their assigned condition. Each assessment included 10 items covering support reactions, equilibrium reasoning, load-position effects, and interpretation of shear force and bending moment relationships. The two forms differed but targeted the same concepts. Items combined multiple-choice and short-response formats to capture both conceptual reasoning and procedural understanding.

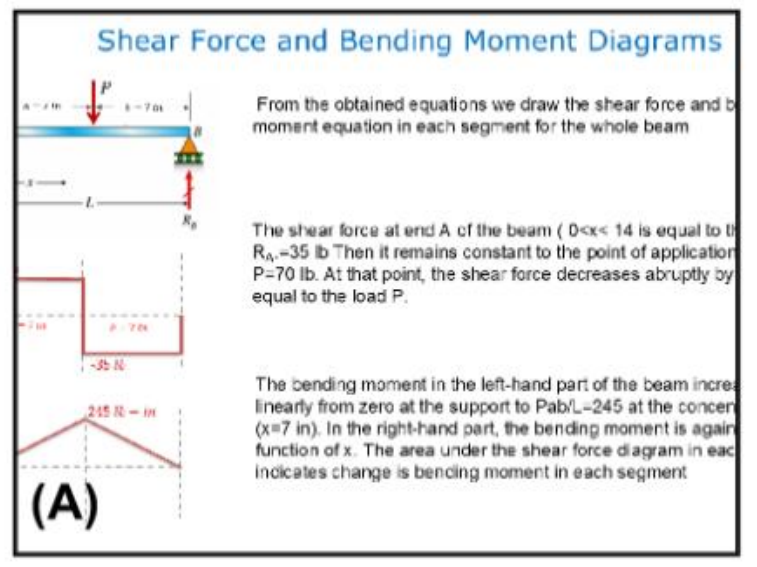

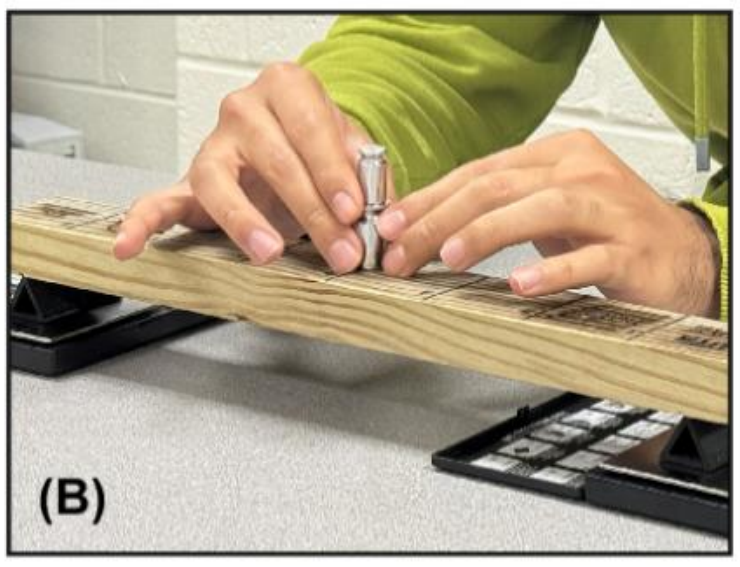

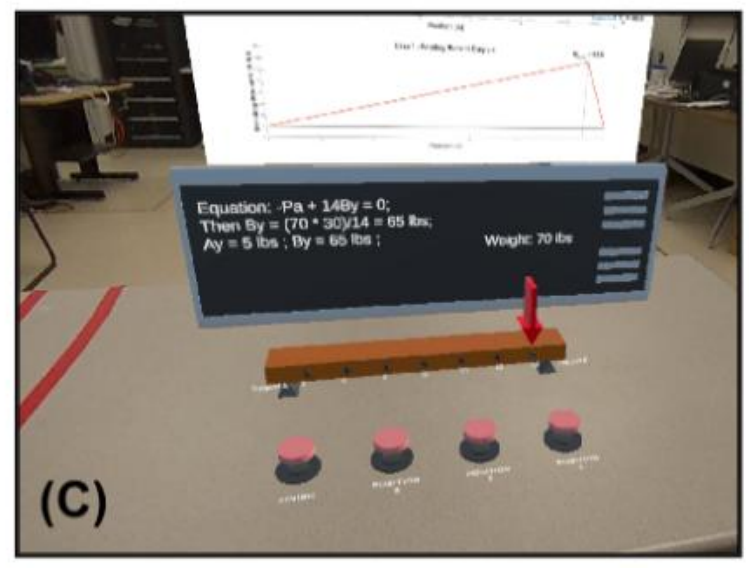

***Figure 2*** *Instructional conditions used in the study. (A) Common lecture-based instruction used to establish a shared baseline across groups. (B) Tangible Engineering Toolkit condition, in which participants manipulated a physical beam setup and observed support reactions. (C) Mixed Reality*

*condition, in which participants interacted with a spatially grounded virtual beam environment and observed corresponding system responses.*

The quizzes were developed for this study and reviewed by subject-matter experts to ensure content alignment and relevance. Short-response items were scored by a single scorer using a predefined rubric and answer key; scoring ambiguity was minimal due to the objective nature of responses. The baseline quiz served as a measure of prior knowledge and as a covariate in the primary analysis, while the post-instruction quiz served as the primary learning outcome. Full instruments are provided in the Appendix.

#### 3.8.2 Usability

Participants in the MR and ET conditions completed the System Usability Scale (SUS), a 10-item instrument scored from 0 to 100 (Bangor et al., 2008; Lewis & Sauro, 2009). Usability was included as a key component of learner experience and a factor influencing the practical adoption of interactive instructional tools.

#### 3.8.3 Learner self-efficacy and experience

Participants in the MR and ET conditions completed a post-instruction questionnaire assessing attitude, perceived skill, comprehension, and agency as dimensions of learner self-efficacy and experience. The instrument included Attitude (3 items), Skill (2 items), Comprehension (4 items), and Agency (5 items), for a total of 14 items, rated on a 5-point Likert scale ranging from 1 (Strongly Disagree) to 5 (Strongly Agree), with negatively worded items reverse coded. The questionnaire was adapted from established constructs and used as an exploratory measure rather than a fully validated scale. These constructs were intended to capture learners' perceptions of engagement, understanding, and confidence during the instructional experience.

#### 3.8.4 Virtual Reality Sickness Questionnaire

Because the MR condition involved immersive participation, participants in that group completed the VRSQ before and after the MR activity to monitor discomfort and support safe participation (Kennedy et al., 1993; Kim et al., 2018).

#### 3.8.5 Open-ended feedback

All participants were invited to provide open-ended comments about aspects of the learning experience they found helpful or challenging. These responses were treated as supplemental contextual data used to support interpretation of the quantitative findings.

***Table 2 Summary of assessment measures.***

| Measure | Purpose | Participants | Notes |
|---|---|---|---|
| Knowledge quizzes | Conceptual understanding | All | Pre (baseline) + post |
| SUS | Usability | MR, ET | 10 items, 0–100 scale |
| Self-efficacy | Confidence, experience | MR, ET | Skill, comprehension, attitude, agency |
| VRSQ | Discomfort monitoring | MR | Pre and post exposure |
| Open-ended responses | Supplemental context | All | Contextual interpretation |

### 3.9 Data analysis

The primary outcome was post-instruction knowledge score. Analysis of covariance (ANCOVA) was used, with instructional condition as the independent variable and baseline knowledge as a covariate. This approach adjusts for prior knowledge differences, improves precision in small samples, and aligns with the design where all participants shared a common lecture baseline but differed in post-lecture experience. ANCOVA assumptions were evaluated, including homogeneity of regression slopes, normality of residuals, and homogeneity of variance. Given the modest sample size, results were also validated using a permutation-based ANCOVA (Freedman-Lane procedure) to reduce reliance on parametric assumptions. Two planned contrasts compared each interactive condition to lecture-only (MR vs. lecture-only; ET vs. lecture-only), with Holm correction applied to control familywise error. Effect sizes were reported as partial $\eta^2$ for ANCOVA and Cohen's d for pairwise comparisons. Normalized learning gain was calculated to describe relative improvement but treated as descriptive due to sensitivity to baseline scores. Within-group pre-post changes were examined using paired t-tests, with Wilcoxon signed-rank tests as a nonparametric check. These analyses were secondary. Open-ended responses were reviewed for recurring patterns related to visualization, interaction, and conceptual clarity, serving to contextualize quantitative findings rather than as a formal qualitative analysis. Analyses were conducted in Python, SAS, and MATLAB.

### 3.10 Ethical considerations

The study was approved by the Institutional Review Board of the authors' institution under a protocol blinded for review. Participants were informed of their right to withdraw without penalty at any time and provided informed consent prior to participation. All data were de-identified prior to analysis and stored on secure institutional infrastructure accessible only to authorized research personnel.

## 4. Results

### 4.1 Instrument reliability

Internal consistency of the learner self-efficacy measure was examined using responses from the two interactive conditions combined ($n = 16$). Reliability varied across subscales. The Comprehension ($\alpha = .845$) and Attitude ($\alpha = .757$) subscales showed good internal consistency, and Agency showed acceptable consistency ($\alpha = .661$). The Confidence composite, formed from the Skill and Comprehension items, also showed acceptable reliability ($\alpha = .715$). In contrast, the two-item Skill subscale produced a negative alpha ($\alpha = -1.29$), indicating that the items did not function consistently as a standalone scale in this sample. This may reflect item misalignment, reverse-wording effects, or instability associated with a small sample and a two-item scale. The Satisfaction composite, formed by combining Attitude and Agency items, showed poor reliability ($\alpha = .128$), suggesting that these items do not operate as a single coherent construct. The full 14-item scale yielded $\alpha = .626$, which is acceptable for an exploratory, context-specific measure but does not support strong interpretation of all subscales. Based on these results, the Confidence composite was retained with caution, while the Satisfaction composite was treated as descriptive only. Given these reliability limitations, self-efficacy results are interpreted as exploratory and are not used to support the primary conclusions of the study. The knowledge assessments were developed for this study and reviewed by a subject-matter expert to support content validity; internal consistency was not estimated.

***Table 3 Internal consistency of learner self-efficacy subscales and composites***

| Scale / subscale | Number of items | Cronbach's α | Interpretation |
|---|---|---|---|
| Attitude | 3 | 0.757 | Good |
| Skill | 2 | -1.29 | Inconsistent, interpret with caution |
| Comprehension | 4 | 0.845 | Good |
| Agency | 5 | 0.661 | Acceptable |
| Confidence composite (Skill + Comprehension) | 6 | 0.715 | Acceptable |

| Scale / subscale | Number of items | Cronbach's α | Interpretation |
|---|---|---|---|
| Satisfaction composite (Attitude + Agency) | 8 | 0.128 | Poor, descriptive only |
| Full self-efficacy scale | 14 | 0.626 | Acceptable |

***Note.*** *Cronbach's alpha values were computed using responses from participants in the Mixed Reality and Engineering Toolkit conditions only (n = 16). Negative alpha values indicate that items did not vary consistently with each other. This can occur when items are poorly aligned, reverse-worded without adjustment, or estimated from very small samples. Based on these results, the Confidence composite was retained with caution, while the Satisfaction composite was treated as descriptive only.*

### 4.2 Baseline equivalence and descriptive statistics

Before testing instructional effects, baseline equivalence was examined to determine whether groups began with similar prior knowledge. This was important because the main analysis compares post-instruction performance across conditions, and pre-existing differences could affect interpretation. A one-way ANOVA on baseline quiz scores showed no significant difference among the Traditional, Mixed Reality (MR), and Engineering Toolkit (ET) groups, $F(2, 21) = 0.44$, $p = 0.650$, indicating comparable starting levels of conceptual understanding. A separate one-way ANOVA on raw post-test scores approached significance, $F(2, 21) = 3.34$, $p = 0.056$, suggesting emerging group differences after instruction. Because small baseline differences can still affect precision in a modest sample, baseline score was retained as a covariate in the primary analysis. Descriptive statistics for baseline, post-test, and raw change scores are reported in Table 4. The Traditional group showed a slight average decline, whereas both interactive conditions improved, with the largest raw mean gain in MR followed by ET.

***Table 4 Descriptive statistics for baseline, post-test, and raw change scores by instructional condition.***

| Group | n | Baseline M (SD) | Post-test M (SD) | Raw change M (SD) |
|---|---|---|---|---|
| Traditional | 8 | 6.38 (1.69) | 6.00 (2.51) | -0.38 (1.85) |
| Mixed Reality | 8 | 6.12 (1.46) | 8.12 (1.46) | 2.00 (1.20) |
| Engineering Toolkit | 8 | 6.88 (1.73) | 8.12 (1.55) | 1.25 (2.65) |

***Note.*** *Raw change scores were computed as post-test minus baseline score and are reported descriptively only. Because the primary comparison of instructional effects adjusts for prior knowledge, inferential conclusions are based on the baseline-adjusted analyses reported in the following sections.*

### 4.3 ANCOVA assumption checks

Before interpreting the primary analysis, ANCOVA assumptions were evaluated because the main test compares post-instruction scores across conditions while adjusting for baseline knowledge. The homogeneity-of-regression-slopes assumption was tested through the interaction between baseline score and instructional condition. This interaction approached but did not reach conventional significance ($p = .055$), indicating that the relationship between baseline knowledge and post-test performance did not differ reliably across groups. Although close to the conventional threshold, this result did not justify rejecting the common-slope assumption, so the standard ANCOVA model was retained. Additional diagnostics supported model adequacy. Residuals were approximately normal according to the KS normality test on residuals ($p > .05$), and adjusted residual variances did not differ significantly across groups according to Levene's test ($p > .05$). Together, these results suggested that ANCOVA assumptions were reasonably satisfied. Because the sample was modest, a permutation-based ANCOVA using the Freedman-Lane procedure was also conducted as a robustness check for the main inferential result.

### 4.4 Primary analysis: baseline-adjusted instructional effects

The primary analysis tested whether post-instruction knowledge scores differed across the three instructional conditions after adjusting for baseline knowledge. ANCOVA was used to compare post-test performance while controlling for pre-instruction differences.

The analysis showed a significant main effect of instructional condition on post-test scores, $F(2, 20) = 3.60$, $p = .048$, partial $\eta^2 = .263$. This indicates that instructional modality explained a meaningful share of variance in post-test performance after baseline knowledge was taken into account. The effect size was large, suggesting that the type of post-lecture learning support was associated with substantively different learning outcomes. Baseline knowledge showed a moderate but non-significant association with post-test performance, $F(1, 20) = 3.20$, $p = .090$, partial $\eta^2 = .137$. A permutation-based ANCOVA using the Freedman-Lane procedure produced a similar result ($p = .050$), supporting the robustness of the finding in this modest sample. Table 5 summarizes the results.

***Table 5 ANCOVA summary for post-test performance.***

| Source | SS | df | F | p | Partial $\eta^2$ |
|---|---|---|---|---|---|
| Teaching modality | 23.3 | 2, 20 | 3.6 | .048* | 0.263 |
| Baseline quiz (covariate) | 10.4 | 1, 20 | 3.2 | 0.09 | 0.137 |
| Error | 65.4 | 20 | | | |

**Note. Post-test knowledge score was the dependent variable. Baseline quiz score was entered as a covariate to adjust for prior knowledge. A permutation-based ANCOVA using the Freedman-Lane procedure with 5,000 permutations yielded a similar result for teaching modality (p = .050), supporting the robustness of the model.**

To clarify the size and direction of the group differences, adjusted post-test means were estimated at the grand-mean baseline score of 6.46. At this common baseline level, the Traditional condition had the lowest adjusted post-test score ($M = 6.04$). Both interactive conditions showed higher adjusted performance: the Engineering Toolkit yielded $M = 7.95$, and Mixed Reality yielded $M = 8.27$. These adjusted means indicate that, after accounting for prior knowledge, both interactive conditions were associated with higher post-instruction performance than lecture-only instruction.

Planned contrasts were used to compare each interactive condition with the Traditional condition. These contrasts were corrected using the Holm procedure to control familywise error across the two focal comparisons. Mixed Reality outperformed Traditional instruction, $t(20) = 2.46$, $p = .046$, $d = 1.24$. Engineering Toolkit also outperformed Traditional instruction, $t(20) = 2.09$, $p = .050$, $d = 1.06$. The comparison between Mixed Reality and Engineering Toolkit was not statistically significant ($p > .05$). Together, these results indicate that the main difference in learning outcomes was between interactive and non-interactive instruction rather than between the two interactive modalities themselves. Adjusted means and planned contrasts are reported in Table 6, and the adjusted group pattern is visualized in Figure 3.

***Table 6 Adjusted post-test means and planned contrasts***

| Condition / comparison | Adjusted mean | 95% CI | Test statistic | p | Effect size |
|---|---|---|---|---|---|
| Traditional | 6.04 | [4.70, 7.37] | | | |
| Engineering Toolkit | 7.95 | [6.60, 9.30] | | | |
| Mixed Reality | 8.27 | [6.93, 9.61] | | | |
| Mixed Reality vs. Traditional | | | **t**(20) = 2.46 | 0.046 | $d = 1.24$ |
| Engineering Toolkit vs. Traditional | | | **t**(20) = 2.09 | 0.05 | $d = 1.06$ |

| Condition / comparison | Adjusted mean | 95% CI | Test statistic | p | Effect size |
|---|---|---|---|---|---|
| Mixed Reality vs. Engineering Toolkit | | | | > .05 | |

**Note.** *Adjusted means were estimated at the grand-mean baseline score of 6.46. Planned contrasts compared each interactive condition with the Traditional condition. Reported p values reflect Holm correction across the two focal contrasts. Cohen's d values indicate large effects for both interactive versus Traditional comparisons.*

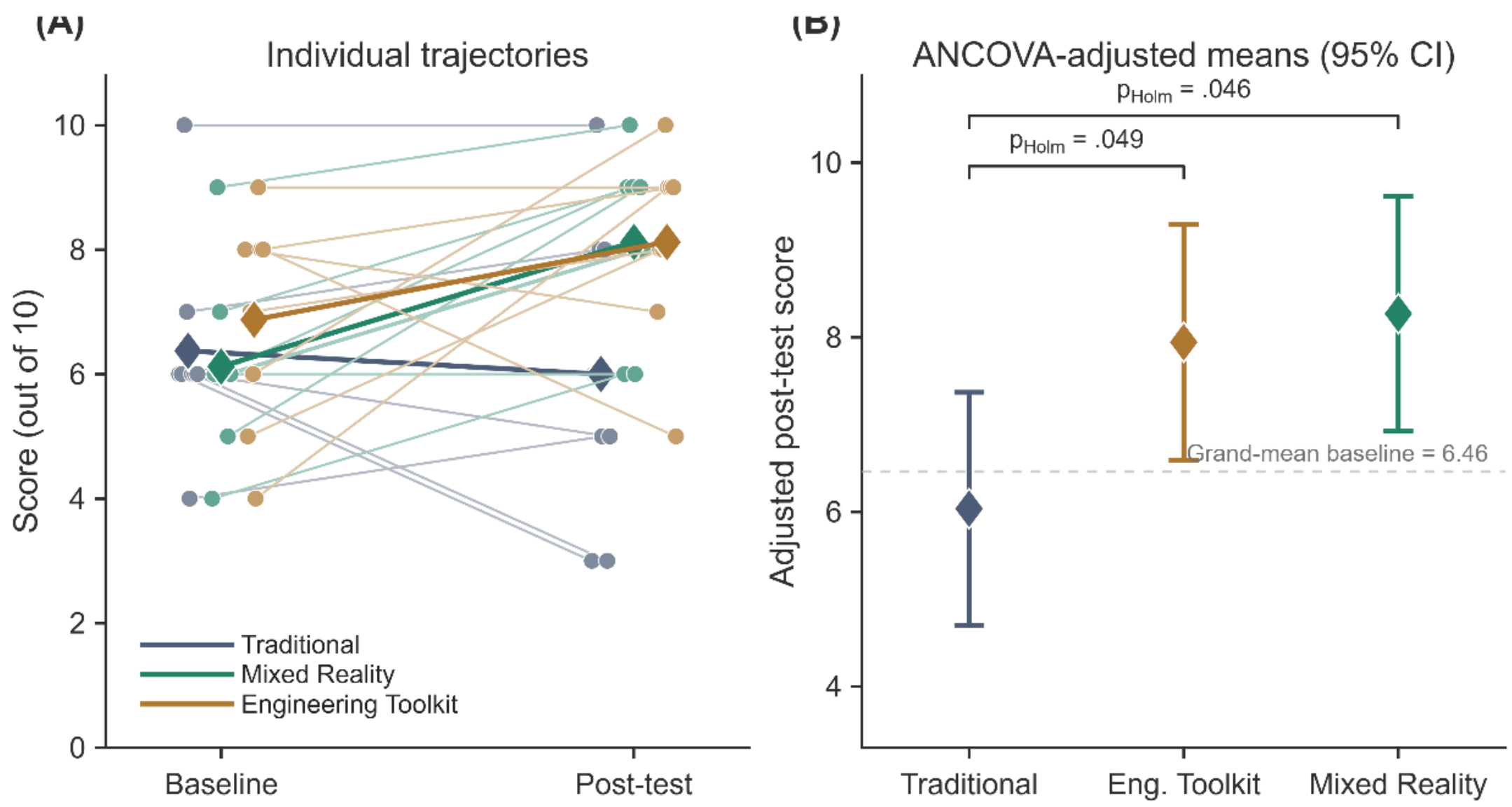

***Figure 3*** *Baseline-adjusted post-test by modality: (A) participant baseline and post-test scores with within-group regression lines; (B) shows adjusted post-test means at grand-mean baseline, 95% confidence intervals, and individual scores.*

### 4.5 Normalized learning gains and within-group change

To complement the baseline-adjusted ANCOVA results, normalized learning gain was calculated as a descriptive measure of improvement relative to starting score. This shows how much each group improved from its own baseline, while the ANCOVA remains the main inferential analysis because it adjusts directly for prior knowledge. Mixed Reality showed the largest normalized gain (M = 0.57, SD = 0.31), indicating strong conceptual improvement. Engineering Toolkit showed a smaller positive gain (M = .16, SD = .82), while the Traditional condition showed no average gain (M = -.11, SD = .50). These results are reported in Table 7 and are broadly consistent with the primary analysis, with the strongest relative improvement in the MR group. These gain values should be interpreted alongside the ANCOVA results, not on their own. Although the ET and MR groups had similar adjusted post-test means, their normalized gains differed because the ET group started with a somewhat higher baseline score and therefore had less room to improve within the normalized-gain formula. For this reason, normalized gain is treated here as a supporting descriptive analysis. Within-group pre-post comparisons provided additional context. Only the MR group showed reliable improvement from baseline to post-test, t(7) = 4.70, p = .002, with the Wilcoxon signed-rank test supporting the same result, p = .016. The Traditional group, t(7) = -0.60, p = .580, and the ET group, t(7) = 1.30, p = .230, did not show reliable improvement. These results are summarized in Table 8 and are reported as supporting analyses only.

***Table 7 Normalized learning gain by instructional modality.***

| Modality | Mean g (SD) | Baseline M (SD) | Interpretation |
|---|---|---|---|
| Mixed Reality | 0.57 (0.31) | 6.12 (1.46) | Strong conceptual gain |
| Engineering Toolkit | 0.16 (0.82) | 6.88 (1.73) | Modest gain |
| Traditional | -0.11 (0.50) | 6.38 (1.69) | No systematic improvement |

**Note.** *Normalized gain was computed as g = (post-test score - baseline score) / (maximum possible score - baseline score). Values are reported descriptively to show relative improvement from each group's own starting point. Because normalized gain is sensitive to baseline score, inferential conclusions in this study are based primarily on the ANCOVA reported above.*

***Table 8 Within-group pre-post comparisons.***

| Group | t | p | Wilcoxon p | Interpretation |
|---|---|---|---|---|
| Traditional | -0.6 | 0.58 | 0.688 | No significant improvement |
| Mixed Reality | 4.7 | 0.002 | .016* | Significant improvement |
| Engineering Toolkit | 1.3 | 0.23 | 0.266 | Small, non-significant improvement |

**Note. Paired-samples t tests and Wilcoxon signed-rank tests were used as complementary checks of within-group change from baseline to post-test. These analyses are secondary and descriptive and are not used in place of the primary between-group ANCOVA.**

### 4.6 Usability and learner self-efficacy

In addition to learning outcomes, the two interactive conditions were compared on usability and learner self-efficacy. These measures were included to provide a broader view of the learning experience, especially because both interventions were designed as post-lecture supports rather than stand-alone instructional replacements.

#### 4.6.1 System usability

Usability was measured using the System Usability Scale (SUS) in the Mixed Reality and Engineering Toolkit conditions. The Engineering Toolkit had a mean SUS score of 86.3 (SD = 9.4), which falls in the excellent range. Mixed Reality had a mean score of 69.1 (SD = 12.2), which falls in the acceptable to good range. An independent-samples t test showed that usability was significantly higher for the Engineering Toolkit than for Mixed Reality, $t(14) = 3.16$, $p = .007$, $d = 1.58$. This suggests that, although both systems were usable, participants found the tangible toolkit easier to use overall. SUS and learner self-efficacy results are reported in Tables 9 and 10.

#### 4.6.2 Learner trait scores

Learner self-efficacy was examined through Attitude, Skill, Comprehension, and Agency. As shown in Table 9, the Engineering Toolkit had higher mean scores than Mixed Reality on Skill and Comprehension. The direction reversed for Attitude, where Mixed Reality showed a slightly higher mean than the Engineering Toolkit (MR M = 2.13, ET M = 1.75), though both values reflect modest engagement overall. Differences on Agency were smaller and in the same direction as Skill and Comprehension. This pattern suggests clearer group differences in perceived competence than in overall engagement, with the exception of Attitude, which favoured MR. Because this measure was exploratory and some subscales had weaker reliability, these scores are best interpreted descriptively.

#### 4.6.3 Confidence and satisfaction composites

To summarize these results, two composites were examined. Confidence, based on Skill and Comprehension, was significantly higher in the Engineering Toolkit group (M = 4.00, SD = .40) than in the Mixed Reality group (M = 3.50, SD = .20), $t(14) = 3.18$, $p = .007$, $d = 1.59$. Satisfaction, based on Attitude and Agency, did not differ significantly between the Engineering Toolkit (M = 3.16, SD = .29) and Mixed Reality (M = 2.92, SD = .34) groups, $t(14) = 1.48$, $p = .160$, $d = .74$. Because Satisfaction showed poor reliability, this result should be treated cautiously. Taken together, these results suggest that the Engineering Toolkit was associated with a stronger sense of competence and ease of use, while the two interactive conditions were more similar in overall satisfaction with the learning experience.

***Table 9 Usability and learner trait scores for the interactive conditions.***

| Measure | Engineering Toolkit M (SD) | Mixed Reality M (SD) |
|---|---|---|
| SUS | 86.3 (9.4) | 69.1 (12.2) |
| Attitude | 1.75 (0.71) | 2.13 (0.74) |
| Skill | 3.00 (0.76) | 2.44 (0.63) |
| Comprehension | 4.67 (0.49) | 3.88 (0.35) |
| Agency | 4.03 (1.08) | 3.65 (0.74) |

**Note.** *SUS was scored on a 0 to 100 scale. Learner trait scores were obtained from the post-instruction self-efficacy measure completed by participants in the Engineering Toolkit and Mixed Reality conditions only. Because the learner self-efficacy measure was exploratory and customized to the present study, these subscale means are reported primarily for descriptive comparison.*

***Table 10 Independent-samples comparisons for usability, confidence, and satisfaction (Engineering Toolkit vs. Mixed Reality).***

| Outcome | Engineering Toolkit M (SD) | Mixed Reality M (SD) | t(14) | p | Cohen's d | 95% CI |
|---|---|---|---|---|---|---|
| SUS | 86.3 (9.4) | 69.1 (12.2) | 3.16 | .007** | 1.58 | [4.8, 29.6] |
| Confidence | 4.00 (0.40) | 3.50 (0.20) | 3.18 | .007** | 1.59 | [0.15, 0.85] |
| Satisfaction† | 3.16 (0.29) | 2.92 (0.34) | 1.48 | 0.16 | 0.74 | [-0.10, 0.58] |

**Note.** Confidence was computed from the Skill and Comprehension items. Satisfaction was computed from the Attitude and Agency items. The Satisfaction comparison should be interpreted cautiously because the composite showed poor internal consistency in this sample. ** $p < .01$. † Descriptive interpretation recommended.

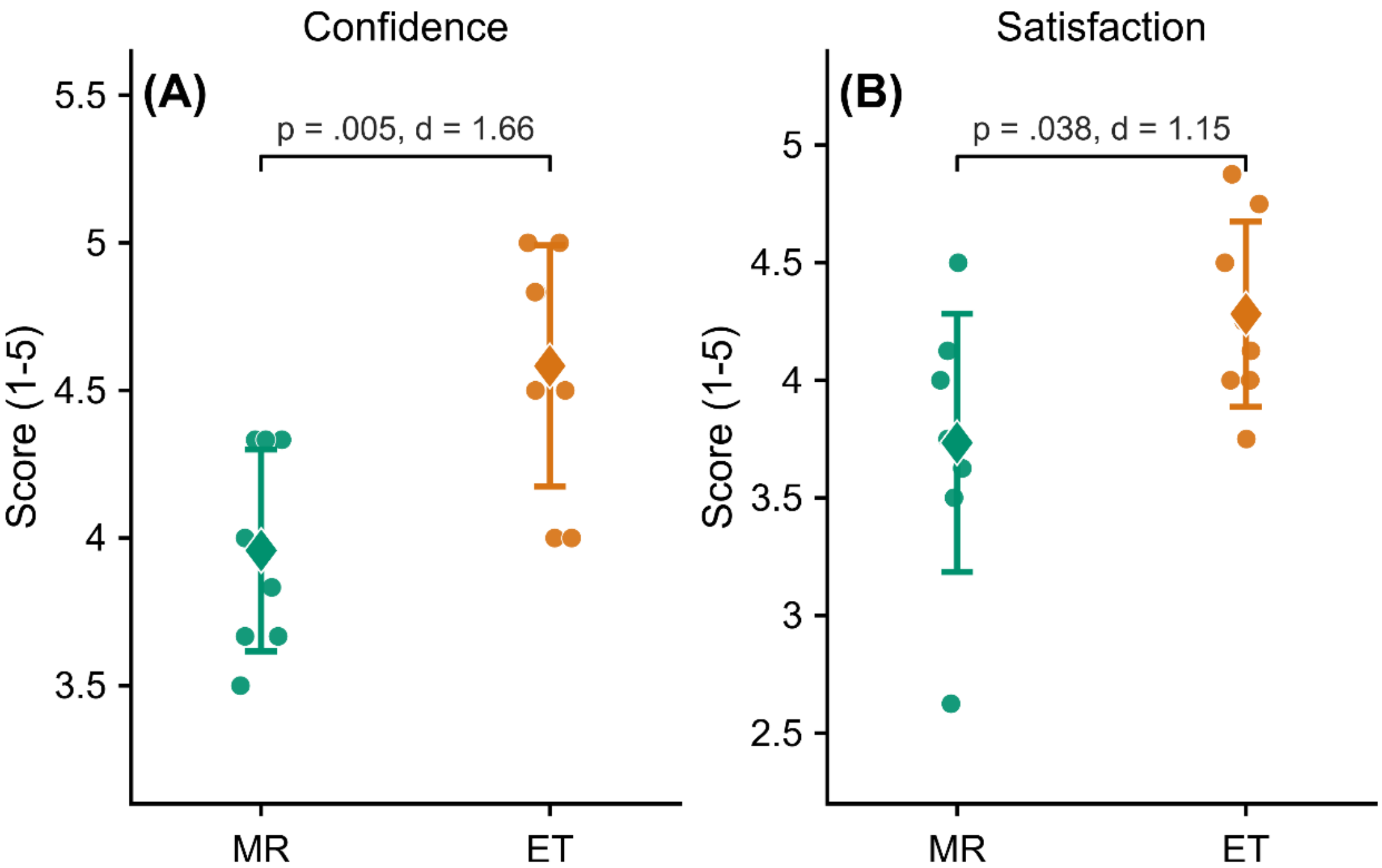

***Figure 4*** *Usability and learner self-efficacy outcomes for interactive conditions. Panel (A) shows the Confidence composite (Skill + Comprehension), Panel (B) shows the Satisfaction composite (Attitude + Agency), and Panel (C) shows System Usability Scale (SUS) scores for the Engineering Toolkit and Mixed Reality conditions. Individual participant scores are shown along with group means and 95% confidence intervals.*

### 1.1. 4.7 Virtual reality sickness

Virtual reality sickness was assessed in the MR condition (n = 8) using the Virtual Reality Sickness Questionnaire (VRSQ; Kim et al., 2018). Pre-session total VRSQ scores were low (M = 4.69, SD = 6.26), and post-session scores were similarly low (M = 4.48, SD = 4.32), with a mean change of −0.21 points. An exact sign-flip permutation test indicated no significant change from pre to post (p = .909). The proportion of participants reporting any symptom was identical at pre and post (4 of 8, 50%), and a McNemar's exact test confirmed no change in symptom prevalence (p = 1.00). Bootstrap 95% confidence intervals for the change score were consistent with no meaningful increase: Total VRSQ 95% CI [−5.36, 5.31].

At the subscale level, Oculomotor scores showed a slight decrease (pre M = 5.73, SD = 8.55; post M = 4.69, SD = 5.35; p = .688), and Disorientation scores remained near floor throughout (pre M = 3.67, SD = 6.92; post M = 4.27, SD = 6.11; p = .844). No participant reported any headache, fullness of head, dizziness with eyes closed, or vertigo at pre-session, and post-session reports were absent or rare. Difficulty focusing showed the largest individual decrease (mean change = −0.38 item units), consistent with oculomotor fatigue resolving after task completion rather than accumulating. Taken together, these findings indicate that the MR session was well-tolerated, with no group-level increase in VR sickness observed. The results are displayed in Figure 5.

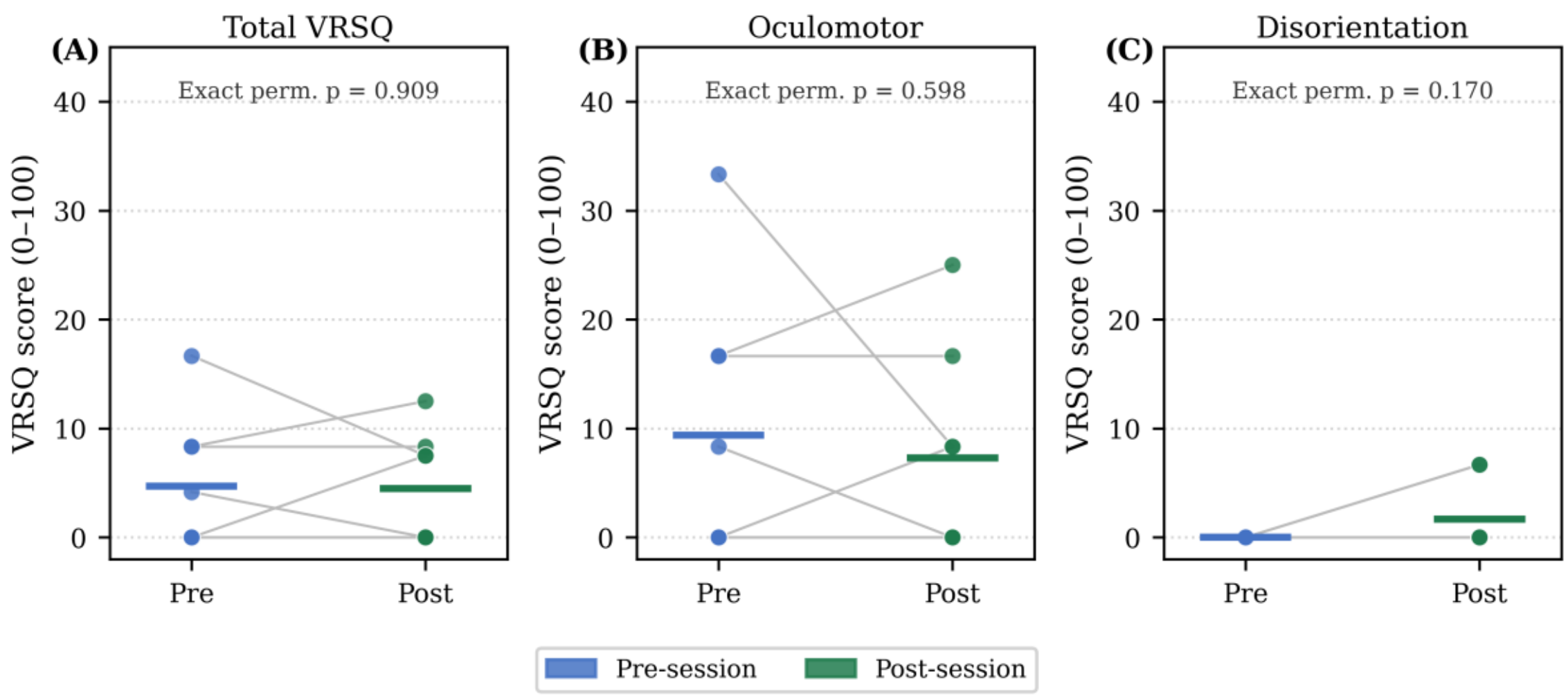

***Figure 5 Individual pre-session and post-session VRSQ scores for the Mixed Reality condition (n = 8). Panels show (A) Total VRSQ, (B) Oculomotor subscale, and (C) Disorientation subscale. Grey lines connect paired observations; filled circles indicate group means. Exact sign-flip permutation p values are shown for each outcome.***

### 4.8 Open-ended feedback

Open-ended responses were reviewed to provide additional context for the quantitative findings. Because these responses were intended to supplement rather than replace the quantitative analyses, they are reported here as exploratory contextual feedback. Three recurring patterns were observed. First, participants often identified visualization as important for understanding beam behavior. Several comments suggested that mechanics concepts became easier to understand when relationships among load position, support reactions, and structural response could be seen more directly. This pattern was consistent with the stronger learning outcomes in the two interactive conditions. Second, participants in the Engineering Toolkit condition often highlighted the value of hands-on interaction. They noted that physically moving the load and observing changes in support reactions made equilibrium relationships more concrete. This aligns with the higher usability and confidence ratings for the toolkit condition. Third, participants in the Mixed Reality condition described the environment as helpful for visualizing otherwise hidden relationships, while also noting that interaction required some initial adjustment. This matches the quantitative results, where MR showed the strongest learning gains but lower usability and confidence ratings than the Engineering Toolkit. Taken together, these responses contextualize the overall quantitative pattern but are not treated as formal qualitative evidence.

## 5. Discussion

This pilot study examined whether two post-lecture interactive interventions, Mixed Reality (MR) and a tangible Engineering Toolkit (ET), could improve conceptual learning in solid mechanics beyond lecture-only instruction. Both interactive conditions were associated with better baseline-adjusted learning outcomes than lecture alone, but they differed across measures. MR showed the strongest relative learning gains, while ET showed higher usability and stronger learner confidence. These results suggest that interactive post-lecture supports are not interchangeable and should be judged by both learning outcomes and learner experience. The findings matter for two reasons. First, interactive tools added value even as lecture extensions rather than replacements. Second, different forms of interactivity appeared to support different aspects of learning. The rest of the discussion interprets these preliminary findings in relation to the research questions, prior work, and instructional design.

### 5.1 Main findings in relation to the research questions

The first research question asked whether MR and the tangible toolkit improved conceptual understanding compared with lecture-only instruction. The results support that conclusion. After accounting for baseline knowledge, instructional modality significantly affected post-test performance, and both interactive conditions outperformed the Traditional condition. This is consistent with prior work showing that interactive representations can support conceptual learning by making abstract relationships easier to inspect, manipulate, and interpret (Kraus et al., 2021; Pàmies-Vilà et al., 2025; Sadowski & Jankowski, 2021). The present study extends that work by showing this effect within a shared post-lecture design, where all participants received the same initial instruction and the learning difference was associated with the added intervention. This matters because many educational settings still use lectures, and the practical question is often how to strengthen learning after lecture rather than replace it.

The second research question asked how learner confidence and satisfaction differed between the two interactive conditions. The pattern was mixed. ET showed higher confidence than MR, while satisfaction did not differ reliably between groups. This suggests that the two modalities supported different aspects of the learning experience. The stronger confidence in ET is consistent with the idea that physical manipulation provides a more direct link between action and outcome, which may strengthen learners' sense of competence. The lack of a clear difference in satisfaction suggests that both modalities were experienced positively overall, although this result should be interpreted cautiously because the satisfaction composite showed weak internal consistency. Even so, the broader pattern is informative: the condition with the strongest learning gain was not the one with the highest confidence rating.

The third research question asked how learners perceived the usability of the two interactive conditions. Both systems were rated as usable, but ET received higher ratings than MR. This is important because usability shapes how easily learners can engage with an instructional tool. The higher usability of ET likely reflects the low interaction overhead of a physical apparatus, where cause and effect are directly visible. The lower, though still acceptable, usability of MR is consistent with prior work showing that immersive systems can provide strong visualization benefits while also introducing interface demands that affect learner experience (Koumpouros, 2024; Wu et al., 2023).

Across the three research questions, the overall pattern is clear. Interactive post-lecture supports improved conceptual learning relative to lecture alone, but immersive and tangible modalities did not support learning in the same way. MR was more strongly associated with conceptual improvement, while ET was more strongly associated with confidence and usability. This pattern supports the broader interpretation developed in the following sections.

An additional finding concerns the comfort and safety of the MR condition. Virtual reality sickness, assessed using the Virtual Reality Sickness Questionnaire (Kim et al., 2018), remained low and did not increase significantly from pre- to post-session. Oculomotor symptoms showed a modest non-significant decrease, while Disorientation scores were negligible across all participants. These results suggest that the MR application was well-tolerated in this sample, which is a practically important consideration for classroom adoption. Although virtual reality sickness was not a pre-specified research question, the finding is relevant to interpreting the MR condition: strong learning gains were achieved without meaningful increases in discomfort, which supports the viability of this MR application as a post-lecture support in spatially demanding engineering courses.

### 5.2 Interactive post-lecture learning as the main instructional contribution

An important finding of this study is not only that the interactive conditions showed stronger performance than lecture-only instruction, but that they did so as post-lecture support rather than as replacements for lectures. This matters because lecture remains the dominant format in many engineering classrooms due to time, curriculum, and class-size constraints. In practice, the question is often not whether lecture should be removed, but how learning can be strengthened after students receive an initial explanation.

The present findings suggest that interactive post-lecture supports can play a useful role in that stage of learning. Rather than framing technology as a replacement for teaching, this approach treats it as a targeted extension of teaching, a distinction that aligns with existing thinking in instructional design (Kirkwood & Price, 2014). Because all participants received the same lecture, the differences that emerged afterward reflect what each post-lecture condition added. This helps reduce confounding from differences in content or instructor delivery, though other factors such as novelty or time-on-task cannot be fully ruled out.

The results suggest that, at least in this context, post-lecture interactivity may be especially helpful in topics where students must move from verbal or diagrammatic explanation to more integrated conceptual understanding. In solid mechanics, this means linking load placement, support reactions, and diagram interpretation to structural behavior. A lecture can introduce these ideas efficiently, but interactive post-lecture activities may help learners test, inspect, and reorganize them after initial exposure (Ainsworth, 2006; Hegarty, 2004b; Streveler et al., 2008). This also helps explain why both interactive conditions outperformed lecture-only instruction despite their different formats. What they shared was not the same interface, but the same instructional function: both let learners revisit the same concepts through action, feedback, and visible response. From this perspective, the present study contributes not just a comparison of tools, but preliminary evidence suggesting that post-lecture learning deserves closer attention as a distinct phase of instructional design in spatially demanding domains.

### 5.3 Complementary strengths of immersive and tangible modalities

A central finding of this study is that the two interactive conditions did not produce the same pattern of outcomes, even though both improved learning relative to lecture-only instruction. Mixed Reality was associated with the strongest conceptual gain, whereas the Engineering Toolkit was associated with higher usability and stronger learner confidence. This suggests that immersive and tangible modalities offer complementary strengths rather than interchangeable versions of the same intervention. One way to interpret this pattern is through the kind of representational support each modality provides. MR appears to have been especially effective in helping learners inspect relationships that are difficult to see in a traditional lecture. By allowing learners to manipulate load position and observe structural response in a spatially grounded environment, MR may have reduced the need for internal mental simulation. In beam bending, where learners must connect diagrams, forces, and physical behavior, this kind of support may be especially useful. The ET condition appears to have supported a different aspect of learning. Its higher usability and confidence ratings suggest that learners experienced the tangible setup as more direct and easier to use. Physical manipulation may have made cause-and-effect relationships clearer, supporting perceived competence during exploration (Davishahl et al., 2020; Wilson, 2002). These findings do not suggest that one modality is simply better than the other. Both were effective relative to lecture alone, but their strengths differed. MR was more strongly associated with conceptual improvement, while ET was more strongly associated with ease of use and confidence. This suggests that instructional value depends not only on whether a tool works, but on what kind of support it provides and which aspect of learning it serves.

### 5.4 Why learning and learner experience did not fully align

An important pattern in this study is that the condition with the strongest learning gain was not the one with the highest usability and confidence ratings. Mixed Reality showed the strongest relative improvement in conceptual learning, while the Engineering Toolkit was easier to use and was associated with higher confidence. This matters because it shows that learning performance and learner experience are not the same thing and should not be treated as interchangeable indicators of instructional value. One possible explanation is that the two modalities supported learning in different ways while placing different demands on learners. MR may have supported conceptual change by helping learners inspect spatial relationships that are otherwise hard to visualize. At the same time, the immersive interface may have introduced extra demands related to navigation, control, or adjustment to a less familiar system. These demands may not have blocked learning, but they may have affected how comfortable or confident learners felt. In contrast, the ET condition offered a more direct and transparent interaction format, which may explain its higher usability and confidence ratings. This pattern is important because it challenges the idea that the most usable or confidence-building tool will always produce the strongest learning gain. Experience measures such as usability and confidence are valuable, but they should not replace direct evidence of learning. Looking at both kinds of outcomes gives a more complete view of instructional effectiveness. In this study, that made it possible to see a more nuanced pattern: MR was more strongly associated with conceptual gain, while ET was more strongly associated with interaction clarity and perceived competence.

### 5.5 Implications for instructional design and broader educational use

These findings have implications for both instructional design and classroom use in settings where lecture remains the main form of instruction. The results suggest that post-lecture interactive supports are most useful when they target concepts that students still find abstract or difficult after initial exposure. In this study, that meant linking load position, support reactions, and diagram interpretation to structural behavior. More broadly, this suggests that interactive activities are most valuable when they are designed as targeted conceptual supports rather than as general engagement tools. The results also show that different modalities may serve different instructional purposes. The stronger learning gains in Mixed Reality suggest that immersive tools may be especially useful when learners need help coordinating spatial and conceptual representations or visualizing relationships that are otherwise hard to see. The Engineering Toolkit results highlight the value of interaction clarity. Its higher usability and confidence ratings suggest that learners benefit when action and response are direct, visible, and easy to interpret. This points to an important design principle: interactive learning tools should offer rich representations without adding unnecessary interaction burden. These findings may also matter beyond solid mechanics. Similar demands appear in statics, structural analysis, dynamics, machine design, and other STEM topics that require learners to coordinate symbolic, visual, and conceptual representations. Although the same pattern may not appear in every context, the broader principle remains that well-designed post-lecture interactive supports can add meaningful value when deeper conceptual understanding is needed. The study also shows that instructional technologies should be evaluated using both learning outcomes and learner experience, since each captures a different part of instructional effectiveness.

### 5.6 Limitations

Several limitations should be considered when interpreting these findings. First, the study involved a small sample from a single institution. Although the randomized design and baseline-adjusted analyses strengthen internal validity, the sample limits statistical power and reduces confidence in generalizing the results to other students, courses, and institutions. The findings should therefore be viewed as preliminary evidence of a meaningful pattern rather than a final estimate of effect size. Second, the study examined learning in a single instructional session. The results therefore speak to immediate conceptual understanding, not long-term retention or transfer. It remains unclear whether the advantages of MR and ET would persist over time or extend to more advanced mechanics contexts. Third, the study evaluated post-lecture support, not full replacement of lecture. The findings show how MR and ET function when added after a shared lecture baseline, not how they would perform as primary instructional methods. Fourth, some effects may reflect the value of added post-lecture engagement in general, not only the specific affordances of each modality. Exact time-on-task during the interactive activities was not formally logged, so this factor cannot be separated from modality effects. Fifth, the learner self-efficacy measure was customized, and not all subscales showed strong reliability. In particular, the Satisfaction composite should be interpreted cautiously. The open-ended responses were also intended only to support interpretation. Taken together, these limits define the scope of the findings without changing the main pattern observed.

## 6. Conclusion

This pilot study examined whether two post-lecture interactive interventions, Mixed Reality and a tangible Engineering Toolkit, could strengthen conceptual learning in solid mechanics beyond lecture-only instruction. Both interactive conditions were associated with stronger learning outcomes than lecture alone after baseline knowledge was taken into account. At the same time, the two modalities did not show the same pattern: Mixed Reality was associated with the strongest conceptual gain, while the Engineering Toolkit was associated with higher usability and stronger learner confidence. These findings suggest that interactive post-lecture supports are not interchangeable. Different modalities may support different aspects of learning, and their value should not be judged from a single outcome alone. In this study, immersive support appeared especially helpful for conceptual improvement, while tangible support provided a more transparent and confidence-building experience. The study also found that immersive MR engagement did not produce meaningful virtual reality sickness symptoms, as confirmed by pre- and post-session VRSQ assessments, providing preliminary evidence that this MR application is well-tolerated in a brief instructional context. Additionally, the study highlights the value of treating post-lecture learning as an important instructional phase. Rather than replacing lecture, interactive tools can extend it by helping learners revisit abstract concepts through action, feedback, and visible response. Despite the study's limits, the findings support a more differentiated view of educational technology in engineering learning and warrant replication with larger samples and longer-term learning measures.

## CRediT authorship contribution statement

Author 1: Conceptualization, Data curation, Formal analysis, Investigation, Validation, Writing – original draft, Writing – review & editing. Author 2: Conceptualization, Data curation, Formal analysis, Validation, Writing – review & editing. Author 3: Data curation, Formal analysis, Methodology, Software, Validation, Visualization, Writing – original draft, Writing – review & editing. Author 4: Investigation, Validation, Writing – review & editing. Author 5: Conceptualization, Resources, Supervision, Validation, Writing – review & editing. Author 6: Conceptualization, Resources, Supervision, Validation, Writing – review & editing.

## Acknowledgements

Acknowledgements have been removed for the purposes of double-anonymized review and will be added in the final manuscript.

## Funding

This research did not receive any specific grant from funding agencies in the public, commercial, or non-profit sectors.

**APPENDIX**
**Appendix:**

**Baseline Knowledge Quiz**

1. **What happens to the shear force when a point load is applied to a beam?**
   - ☐ It remains unchanged
   - ☐ It increases or decreases suddenly
   - ☐ It changes gradually along the length
   - ☐ It becomes zero
2. **In which case does the bending moment remain constant between two points?**
   - ☐ When a point load is applied between the points
   - ☐ When a uniformly varying load is present
   - ☐ When no external loads or reactions exist between the points
   - ☐ When a distributed load is present
3. **Where is the maximum bending moment typically located in a simply supported beam with a single central point load?**
   - ☐ At the supports
   - ☐ At the midpoint of the beam
   - ☐ Near the support but not at them
   - ☐ It depends on the length of the beam
4. **If the shear force is constant in a region of a beam, what does this imply about the bending moment?**
   - ☐ It is also constant
   - ☐ It varies linearly
   - ☐ It varies parabolically
   - ☐ It suddenly increases or decreases
5. **What effect does a roller support have on a beam's shear force and bending moment?**
   - ☐ It resists both vertical and horizontal forces
   - ☐ It allows horizontal movement but resists vertical forces
   - ☐ It resists moments but allows rotation

- It prevents bending moments at the support

6. **How does a pin support differ from a fixed support in terms of bending moment resistance?**
    - A pin support resists bending moments, while a fixed support does not
    - A fixed support resists bending moments, while a pin support does not
    - Both resist bending moments equally
    - Neither resists bending moments
    - A triangular shape
7. **What happens to the bending moment at an internal hinge in a beam?**
    - It is maximum
    - It is zero
    - It depends on the applied load
    - It has no significance
8. **If the bending moment at a point is zero, what does this imply about the beam at that point?**
    - There is no force acting at that point
    - The beam is in pure shear at that point
    - The beam experiences no bending stress at that point
    - The beam is at risk of failure
9. **What is the primary purpose of a shear force and bending moment diagram?**
    - To determine the material properties of a beam
    - To analyze the internal load distribution within the beam
    - To calculate the weight of the beam
    - To determine the manufacturing process of the beam
10. **If two beams have identical loading conditions but different boundary conditions, how will their bending moment diagrams compare?**
    - They will be identical
    - The beam with more restraints will have higher bending moments
    - The beam with fewer restraints will have higher bending moments
    - Boundary conditions do not affect bending moments

**Knowledge Quiz**

1. What types of movements are allowed at support B? Select all that apply.

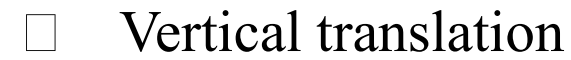

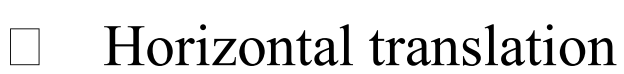

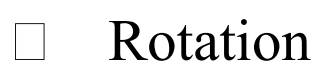

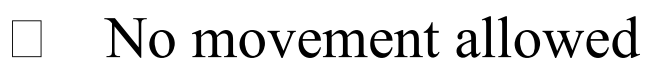

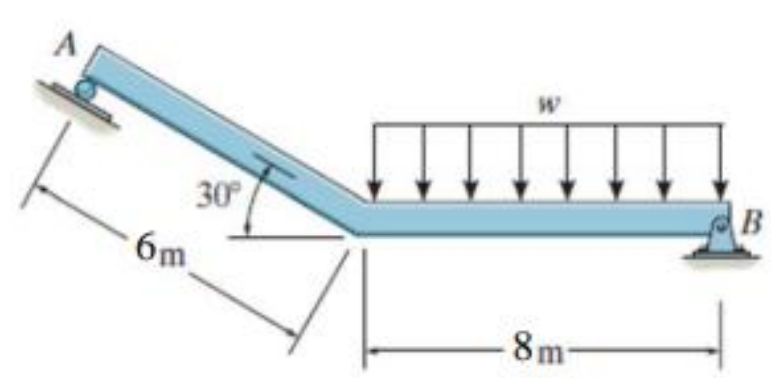

2. In the shear force diagram below what will be the bending moment at point B

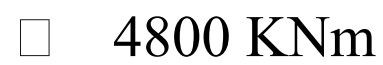

- 2400 KNm
- -1200 KNm
- None

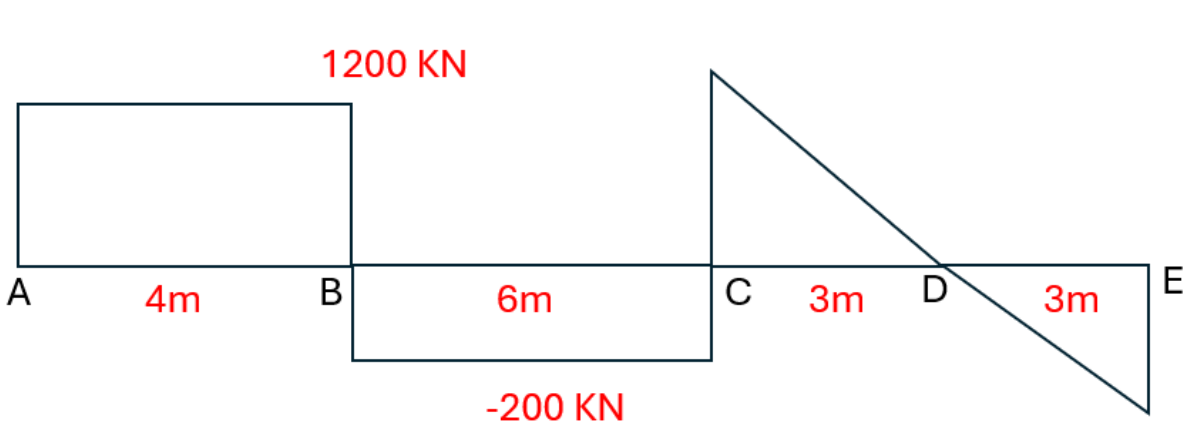

3. For a loading condition of a cantilever beam shown in the figure below, fibers above the neutral axis will be in

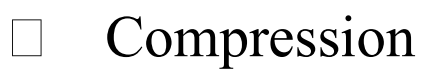

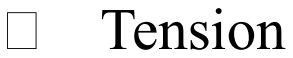

- Will depend on the applied load
- It will vary along the length of the beam

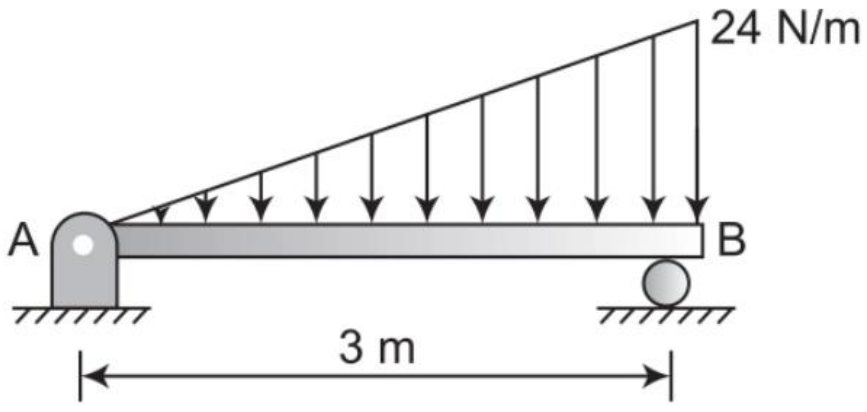

4. For the simply supported beam below, what will be the equivalent point load acting on the beam for the uniformly varying load?

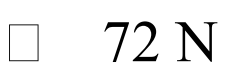

- 12 N
- 36 N
- 16 N

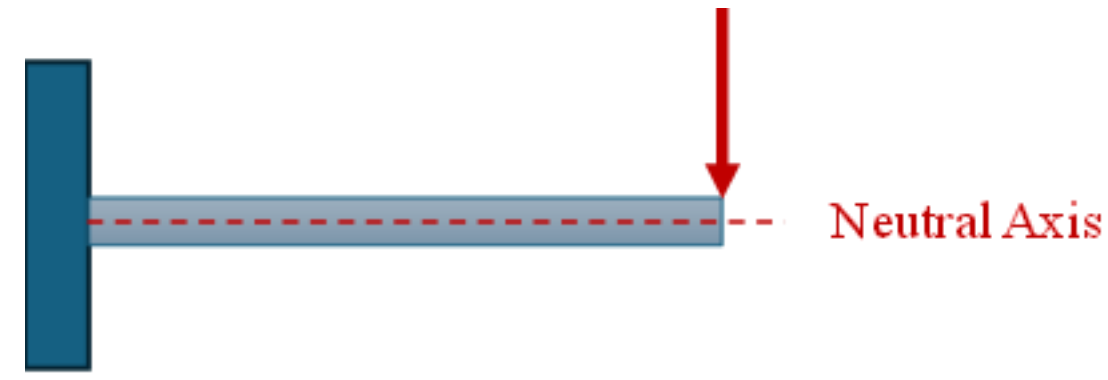

5. What does the shear force represent in a beam?

- ☐ The internal moment resisting bending
- ☐ The internal force resisting transverse loading
- ☐ The reaction force at the support
- ☐ The axial force in the beam

6. In a simply supported beam, how does the bending moment vary at the supports?

- ☐ It is maximum
- ☐ It is zero
- ☐ It depends on the load type
- ☐ It is equal to the reaction force

7. What effect does a roller support have on a beam's shear force and bending moment?

- ☐ It resists both vertical and horizontal forces
- ☐ It allows horizontal movement but resists vertical forces
- ☐ It resists moments but allows rotation
- ☐ It prevents bending moments at the support

8. If you were to analyze support reactions for the shelf of stack of books, what type of loading condition would you consider?

- ☐ Concentrated load
- ☐ Uniformly varying load
- ☐ Uniformly distributed load
- ☐ Concentrated Moment

9. In the given beam setup, why does the reaction at support A exist, while there is no horizontal reaction at support B?

- ☐ Because the roller support at A can resist only vertical forces, while the sliding support at B allows horizontal movement.
- ☐ Because the load distribution is triangular, resulting in a reaction only at A.
- ☐ Because support A is fixed, and fixed support always generates vertical reactions.
- ☐ Because support B does not experience any force in the horizontal direction, it does not develop a reaction.

10. In this cantilever beam, where do you expect to get the maximum bending moment?

- ☐ Point A
- ☐ Point B
- ☐ Point C
- ☐ 2/3 m away from point A